\documentclass[12pt]{article}
\pdfoutput=1
\usepackage[margin=0.8in]{geometry}
\usepackage{amsmath,amssymb,latexsym}
\usepackage{graphicx} 
\usepackage{booktabs}
\usepackage{hyperref}
\usepackage[numbers,sort&compress]{natbib}
\usepackage[toc,page]{appendix}

\usepackage{amsthm}

\usepackage{authblk}

\usepackage{wrapfig}

\usepackage{bbm}

\newcommand{\bR}{\ensuremath{\mathbf{R}}}

\newcommand{\bp}{\ensuremath{\mathbf{p}}}

\newcommand{\bz}{\ensuremath{\mathbf{z}}}
\newcommand{\bH}{\ensuremath{\mathbf{H}}}

\newcommand{\bZ}{\ensuremath{\mathbf{Z}}}

\newcommand{\bbR}{\ensuremath{\mathbb{R}}}

\graphicspath{{figs/}}

\title{One-class Recommendation Systems  with  the Hinge Pairwise Distance Loss and Orthogonal Representations}

\author[ \hspace{-1ex}]{Ramin Raziperchikolaei}
\author[ \hspace{-1ex}]{Young-joo Chung}
\affil[ ]{Rakuten Group, Inc.}
\affil[ ]{\textit {\{ramin.raziperchikola,youngjoo.chung\}@rakuten.com}}
\date{} 

\hypersetup{
pdftitle={One-class Recommendation Systems  with  the Hinge Pairwise Distance Loss and Orthogonal Representations},
pdfauthor={Ramin Raziperchikolaei, Young-joo Chung},
pdfkeywords={One-class Recommendation Systems  with  the Hinge Pairwise Distance Loss and Orthogonal Representations}
}

\begin{document}

\maketitle
\begin{abstract}
In one-class recommendation systems, the goal is to learn a model from a small set of interacted users and items and then identify the positively-related user-item pairs among a large number of pairs with unknown interactions. Most previous loss functions rely on dissimilar pairs of users and items, which are selected from the ones with unknown interactions, to obtain better prediction performance. This strategy introduces several challenges such as increasing training time and hurting the performance by picking "similar pairs with the unknown interactions" as dissimilar pairs. In this paper, the goal is to only use the similar set to train the models.  We point out three trivial solutions that the models converge to when they are trained only on similar pairs: collapsed, partially collapsed, and shrinking solutions. We propose two terms that can be added to the objective functions in the literature to avoid these solutions. The first one is a hinge pairwise distance loss that avoids the shrinking and collapsed solutions by keeping the average pairwise distance of all the representations greater than a margin. The second one is an orthogonality term that minimizes the correlation between the dimensions of the representations and avoids the partially collapsed solution. We conduct experiments on a variety of tasks on public and real-world datasets. The results show that our approach using only similar pairs outperforms state-of-the-art methods using similar pairs and a large number of dissimilar pairs.
\end{abstract}

\section{Introduction}
\label{s:intro}
Feedback in recommendation systems (RSs) can be explicit or implicit. In the case of explicit feedback \cite{Ricci11,Takacs08,Zhoua08,Koren09}, users provide rating values after interacting with the items, which shows how satisfied they are. In the case of implicit feedback  \cite{Ricci11,Hu08,Pan08}, which is a common scenario in the real world, we only know if the user interacted with an item or not (such as clicked, purchased, etc.). The goal of one-class recommendation systems is to solve the implicit feedback prediction problem. It's called a one-class problem since a "no interaction" between a user and an item in the training set does not necessarily mean the user does not like that item. It just means we do not have enough information about their interaction. That's because the set of the items in RSs is huge and the users can't see all the items, i.e., users can only see a small subset of the items and then interact with a few of them.

There are three main steps in learning a RS model to predict the implicit feedback. The first step is to learn user and item representations. This can be done by learning user and item embedding matrices from their IDs \cite{Takacs08,Zhoua08,Koren09,He17,He18}, learning user and item multi-layer perceptrons (MLPs) from the interaction vectors \cite{Sedhain15,Xue17,Dong19} or side information  \cite{Strub16,Zhang17,Dong17}, or learning graph neural networks from the bipartite user-item graph \cite{Berg18,Ying18,Zhang19a,He20}.

The second step is to model the interaction score from the user and item representations. The common functions are dot product \cite{Hu08,Pan08,Takacs08,Koren09,Dong17,Zhang19a,He20,Raziperchikolaei21a}, cosine similarity \cite{Xue17,Ying18}, and neural networks over the concatenated user and item representations  \cite{He17,He18,Dong19,Raziperchikolaei21,Raziperchikolaei22}.

The third step is to optimize a loss function, which gets smaller values as the model gives a larger score to the similar pairs of the user and items compared to the dissimilar ones. Different types of loss functions have been used in the literature. Mean squared error (MSE) loss \cite{Hu08,Pan08,Takacs08,Koren09,Dong17,Zhang19a,Raziperchikolaei21,Raziperchikolaei21a} and Binary cross-entropy (BCE)  \cite{Xue17,He17,Dong19} directly minimize the difference between the predicted and the actual scores. The Bayesian personalized rank (BPR) loss \cite{Rendle09,He18,He20} tries to make the interaction score of the similar pairs greater than the dissimilar ones, instead of directly mapping them to the actual scores. The contrastive learning loss tries to put representations of the similar pairs close to each other and put the dissimilar ones far away \cite{Raziperchikolaei21a,Lin22,Wu19}.

All the above loss functions need both similar and dissimilar pairs of the users and items to learn a model. If we train these loss functions only using the similar pairs, we get a collapsed solution: all representations will be mapped to the same point in the latent space and the model predicts the same interaction score for all the pairs. The performance of the collapsed solution is as bad as assigning random representations to the users and items. So the dissimilar sets are essential in training RS models to avoid the collapsed solution.

In one-class recommendation systems, we only have access to the implicit (known) interactions, and the rest of the interactions are unknown. To create a dissimilar set of users and items, the common approach is to randomly select a set of user and item pairs with the unknown interactions and consider them dissimilar \cite{Xue17,He17,Dong19,Dong17,Zhang19a,Raziperchikolaei21a}. Another strategy is to find out the hard-negatives: the pairs with the unknown interactions that the model has difficulty with classifying as dissimilar \cite{Zhang13,Ding19, Ding20}.

Creating dissimilar pairs from unknown ones is problematic for two main reasons. First, as we show in experiments, we might need a large number of dissimilar pairs to achieve reasonable results, which makes the training slow. Second, using a large number of dissimilar pairs increases the chance of converting a "similar pair with an unknown interaction" to a dissimilar pair, which hurts the performance. This issue is more severe in the case of the hard negative approach, since "similar pairs with unknown interactions" are by definition difficult to be classified as dissimilar, and will be mistakenly taken as hard negatives.

The main goal of this paper is to propose a new objective function that avoids the collapsed solution without using the dissimilar pairs. As we define it mathematically in Section~\ref{s:proposed}, at the collapsed solution, the average pairwise distance between all the user-user, item-item, and user-item representations is $0$. To avoid the collapsed solution, we propose a hinge pairwise distance loss which keeps the pairwise distance above a margin. We show that this new loss function alone is not enough to generate high-quality representations, as the solution might converge to a "partially collapsed solution", where all users and items converge to exactly two unique representations. To avoid this solution, which performs as poorly as the collapsed solution, we propose to minimize the correlation between the dimensions of representations by making them orthogonal.

We conduct extensive experiments on several real-world and public datasets in both warm-start and cold-start settings. The experimental results in Section~\ref{s:exp} show that our approach, which only uses similar pairs, has a better prediction performance than state-of-the-art methods, which use both similar and dissimilar pairs.

\paragraph{Notations.}  We denote the sparse interaction matrix by $\bR \in \mathbb{R}^{m \times n}$, where $m$ and $n$ are the number of users and items, respectively, $R_{jk}>0$ is the interaction value of the user $j$ on the item $k$, and $R_{jk}=0$ means the interaction is unknown.  The goal is to predict the unknown interactions in $\bR$. The $i$th row of a matrix $\bH$ is shown by $\bH_{i,:}$ and the $j$th column is shown by $\bH_{:,j}$. The $d$-dimensional representations of the all users and all items are denoted by $\bZ^u \in \bbR^{m\times d}$ and $\bZ^i \in \bbR^{n\times d}$, respectively. The representation of the $j$th user and $k$th item are denoted by $\bz_{j}^u = \bZ_{j,:}^u$ and $\bz_{k}^i = \bZ_{k,:}^i$, respectively.

\section{Related works}
\label{s:related}
Since our main contribution is proposing a new objective function, we review the different loss functions used in the RS literature. Most of these loss functions are defined based on the predicted user-item interactions from the representations. The most common mappings from representations to predicted interactions are dot product \cite{Hu08,Pan08,Takacs08,Koren09,Dong17,Zhang19a,He20,Raziperchikolaei21a}, cosine similarity \cite{Xue17,Ying18}, and neural networks \cite{He17,He18,Dong19,Raziperchikolaei21,Raziperchikolaei22}:
\begin{equation}
\label{e:map}
\text{dot product:} \hat{R}_{jk} =(\bz_{j}^u)^T \bz_{k}^i, \quad 
\text{cosine:} \hat{R}_{jk} =\frac{(\bz_{j}^u)^T
\bz_{k}^i}{||\bz_{j}^u||||\bz_{k}^i||},  \quad
\text{NNs:} \hat{R}_{jk} =h([\bz_{j}^u,\bz_{k}^i])
\end{equation}
where $[\cdot,\cdot]$ merges the two vectors and $h()$ is a neural network. In \cite{He17,Dong19},$[\cdot,\cdot]$ is concatenation and $h()$ is an MLP. In \cite{He18}, $[\cdot,\cdot]$ is outer product and $h()$ is a convolutional network.

To define the loss functions in RSs, previous works use sets of similar and dissimilar pairs of the users and items. The similar set $S^+$ contains all the users and items that interacted with each other, i.e., $(j,k) \in S^+$ if  $R_{jk}=1$. The dissimilar set $S^-$ contains a subset of the users and items with unknown interactions, i.e., $R_{jk}=0$.

Two popular loss functions in the literature are mean squared error (MSE) \cite{Hu08,Pan08,Takacs08,Koren09,Dong17,Zhang19a,Raziperchikolaei21,Raziperchikolaei21a} and binary cross-entropy (BCE) \cite{Xue17,He17,Dong19}, which directly minimize the difference between the predicted and actual interactions:
\begin{align}
\label{e:rb_loss}
	&l_{\text{BCE}}(\bz_{j}^u,\bz_{k}^i) = - (R_{jk} \log\hat{R}_{jk} + (1-R_{jk}) \log(1-\hat{R}_{jk}) ), \nonumber\\
	&l_{\text{MSE}}(\bz_{j}^u,\bz_{k}^i) = (\hat{R}_{jk}-R_{jk})^2, 
\end{align}
where $j,k \in S^+\cup S^- $.  On the other hand, the BPR loss  \cite{Rendle09,He18,He20} is defined based on the difference between the predicted interactions of the similar and dissimilar pairs:
\begin{equation}
\label{e:bpr_l}
	l_{\text{BPR}}(\bz_{j}^u,\bz_{k}^i)=  - \ln{\sigma(\hat{R}_{jk}-\hat{R}_{jl})},
	\end{equation}
where $j,k \in S^+$ and $j,l \in S^-$, i.e., user $j$ is similar to item $k$ and dissimilar to item $l$. Note that the above loss functions are optimized over the user and item representations, $\bz_{j}^u$ and $\bz_{k}^i$, which are used to generate the predicted interaction. 

Contrastive loss functions \cite{Hadsella06} are also used in recommendation systems \cite{Raziperchikolaei21a,Lin22,Wu19}. The idea is to map representations of similar pairs of the users and items close to each other and the dissimilar ones far away:
\begin{equation}
\label{e:con}
	 l_{\text{contrastive}}(\bz_{j}^u,\bz_{k}^i) = R_{jk} || \bz^{u}_j - \bz^{i}_k||^2 + (1-R_{jk}) \lambda \max{(0,m_d - ||\bz^{u}_j  -  \bz^{i}_k)||)}^2,
\end{equation}
where $m_d$ is a margin and $j,k \in S^+\cup S^- $.

As we explain in the next section, without the dissimilar sets, all the above loss functions will converge to a collapsed solution, which performs as bad as a random assignment of the representations. The goal of this paper is to introduce new terms to let the above loss functions be trained only on similar sets and still avoid the collapse solution.

\section{Proposed method}
\label{s:proposed}
As mentioned in the introduction, there are several issues in using the dissimilar pairs in one-class RSs. First, we are never sure whether the selected pairs with the unknown interactions are dissimilar or not. A "no interaction" does not necessarily mean that the user did not like the item. Selecting similar pairs and using them as dissimilar pairs can hurt the performance. Second, as we show in experiments, previous works use a large number of dissimilar pairs to get reasonable performance, which will increase the training time. Our goal is to learn a recommendation system model by discarding the dissimilar set of users and items, training only on the similar pairs, avoiding the collapsed solution, and achieving state-of-the-art results.

In this section, we first explain how the previous objective functions converge to a collapsed solution when they only use similar pairs. Then, we explain our proposed method by considering the $E_{\text{cont}}(\bZ^u,\bZ^i)$ in Eq.~\eqref{e:con_reg} as the main term of the objective function. We show how we avoid the collapsed, partially collapsed, and shrinking solutions by proposing new terms. Then, we will explain how our objective function works with other loss functions. Finally, we go over the computational complexity of the proposed objective function.

\subsection{The collapsed solution without the dissimilar set}
The loss functions introduced in Section~\ref{s:related} use both similar and dissimilar pairs to predict the actual representations. Let us explain what happens if we only use the similar pairs and discard the dissimilar ones in two cases: 1) using the loss function $l_{\text{MSE}}$ in Eq.~\eqref{e:rb_loss} and dot product mapping function in Eq.~\eqref{e:map}, and 2) using the constrastive loss defined in Eq.~\eqref{e:con}:
\begin{equation}
\begin{split}
\label{e:con_reg}
&E_{\text{MSE}}(\bZ^u,\bZ^i) = \sum_{j,k \in S^+}  ((\bz^{u}_j)^T \bz^{i}_k -1)^2, \\ 
&E_{\text{cont}}(\bZ^u,\bZ^i) = \sum_{j,k \in S^+} || \bz^{u}_j - \bz^{i}_k||^2.
\end{split}
\end{equation}
Both objective functions above are defined on similar pairs $S^+$ and optimizing them will lead to a collapsed solution, where all the users and item representations are mapped to a certain point in the latent space. The optimal solution of the $E_\text{MSE}$ is achieved by mapping all the user and item representations to any $d$-dimensional vector with a unit L2 norm. That's because the dot product of all pairs becomes $1$, 
 i.e., $\hat{R}_{jk} =  (\bz^{u}_j)^T \bz^{i} = 1$, which makes the loss value $0$ for all the terms. In the case of $ E_{\text{cont}}$, the optimal (and collapsed) solution is achieved by mapping all the user and item representations to any $d$-dimensional vector.

At the collapsed solution, which is the result of removing dissimilar pairs from the loss functions, the model always returns the same prediction, no matter what the input pairs are. This works as poorly as a random model. 

Note that different combinations of mapping functions and loss functions can give us different objective functions. In all cases, without dissimilar pairs, the result is a collapsed solution.

\subsection{Avoiding the collapsed solution: hinge pairwise distance loss}
Let us assume the $d$-dimensional representations of the users and items is denoted by $\bZ^u \in \bbR^{m\times d}$ and $\bZ^i \in \bbR^{n\times d}$, respectively. The joint user-item representation is achieved by vertically concatenating the user and item representations, $\bZ = [\bZ^u,\bZ^i] \in \bbR^{(m+n)\times d}$. The pairwise distance between all the representations in $\bZ$ is computed as:
\begin{equation}
\label{e:d_p}
d_{p} = E_{\text{cont}}(\bZ,\bZ) = \frac{1}{(m+n)^2} \sum_{l,s=1}^{m+n} ||\bz_{l} - \bz_s||^2.
\end{equation}
Note that $d_p$ computes distance between all the user-user, item-item, and user-item representations, which is different from $E_{\text{cont}}(\bZ^u,\bZ^i)$ that computes the distance between similar pairs of the users and items. 

Mathematically, at the collapsed solution, we have $d_p=0$. To avoid the collapsed solution, we propose a hinge pairwise distance loss that keeps the average pairwise distance $d_p$ greater than a margin $m_p$. The new objective function can be written as:
\begin{equation}
\label{e:obj1}
E = E_{\text{cont}}(\bZ^u,\bZ^i) + E_{d_p}(\bZ) =  \sum_{j,k \in S^+} || \bz^{u}_j - \bz^{i}_k||^2 + \max{(0,m_p - d_p)}^2.
\end{equation}
Note that $d_p$ involves computing the distances between all the pairs, which could be very time-consuming. As we show below, $d_p$ is equivalent to the summation of the variance of each dimension, which can be computed way faster.

Let us denote the $q$th dimension of the $l$th representation as $z_{l,q}$, and the pairwise distance of the $q$th dimension as $d_p^{q}$.  Then, $d_p$ in Eq.~\eqref{e:d_p} can be separated over the $d$ dimensions:
\begin{equation}
d_p = \sum_{q=1}^d d_p^{q} = \frac{1}{(m+n)^2} \sum_{q=1}^d\sum_{l,s=1}^{m+n} (z_{lq} - z_{sq})^2.
\end{equation}
We can rewrite $d_p^{q}$ as:
\begin{multline}
\label{e:var}
d_p^{q}=  \frac{1}{(m+n)^2} \sum_{l,s=1}^{m+n} (z_{lq} - z_{sq})^2 =  \frac{1}{(m+n)^2}\sum_{l,s=1}^{m+n}  z_{lq}{^2} + z_{sq}{^2} - 2z_{lq}z_{sq} = \\ \frac{2}{(m+n)^2}\sum_{l,s=1}^{m+n}  z_{lq}{^2}  - \frac{2}{(m+n)^2}\sum_{l=1}^{m+n}z_{lq} \sum_{s=1}^{m+n}z_{sq} =  \frac{2}{(m+n)}\sum_{l=1}^{m+n}  z_{lq}{^2} - 2\bar{\bZ}_{:,q}^2 = 2 \text{var}(\bZ_{:,q}).
\end{multline}
So twice the variance of a dimension is equivalent to the pairwise distances of the representations in that dimension. This means that at the collapse solution, the variance of each dimension is $0$, and to avoid this solution, we just need the summation of the variance of the dimensions to be greater than a margin. At the end of this section, we explain how the new formulation helps the computational complexity.

\subsection{Avoiding the partially collapsed solution: orthogonality term}
While the objective function of Eq.~\eqref{e:obj1} avoids the collapsed solution, it might give us low-quality representations by converging to a "partially collapsed solution". This solution returns only two representations for the whole set of users and items: a portion of the users and items are mapped to one representation and the rest of them are mapped to a second representation. Below, we explain when the objective function converges to this solution and how we avoid it by proposing an orthogonality term.

To explain the partially collapsed solution, we consider the user-item bipartite graph. This graph contains two disjoint sets of nodes: set $I$ contains the items and set $U$ contains the users. The edges are determined by $S^+$, i.e., there is an edge between nodes of user $j$ and item $k$ if $(j,k) \in S^+$. Now, let us assume the graph can be partitioned into $V>1$ disjoint components $\{C_v\}_{v=1}^V$, where the edges in each component are denoted by $S_v^{+}$. Since it's a disjoint partition, each node from $U$ and $I$ appears in exactly one component, $\cup\{S_v^{+}\}=S^+$, and there is no edge across the nodes of the two different components. The interaction matrix is more than $99\%$ sparse in real world scenarios, which can easily lead to having multiple disjoint partitions.

The optimal solution of our objective function in the setting above can be achieved by assigning representation $\bp_1$ to all the nodes in some of the components and $\bp_2$ to all the nodes in the rest of the components.  

First, let us show how the above solution makes makes $E_{\text{cont}}=0$. Here, for simplicity, let us assign $\bp_1$ to the components $1,\dots,\frac{v}{2}$ and assign $\bp_2$ to the components $\frac{v}{2}+1,\dots,V$. Then we have:
\begin{multline}
E_{\text{con}} =  \sum_{j,k \in S^+} || \bz^{u}_j - \bz^{i}_k||^2 = \sum_{v=1}^V \sum_{j,k \in S_v^{+}} || \bz^{u}_j - \bz^{i}_k||^2 = \\ \sum_{v=1}^{\frac{V}{2}} \sum_{j,k \in S_v^{+}} || \bz^{u}_j - \bz^{i}_k||^2 + \sum_{v={\frac{V}{2}+1}}^V \sum_{j,k \in S_v^{+}} || \bz^{u}_j - \bz^{i}_k||^2  = \\ \sum_{v=1}^{\frac{V}{2}} \sum_{j,k \in S_v^{+}} ||\bp_1 - \bp_1||^2 + \sum_{v={\frac{V}{2}+1}}^V \sum_{j,k \in S_v^{+}} || \bp_2 - \bp_2||^2=0.
\end{multline}
We can easily prove that $E_{\text{con}}$ becomes $0$ for the other ways of assigning the two representations to the components.

To make the hinge pairwise distance loss $E_{d_p}=0$, we just need to put the two points $\bp_1$ and $\bp_2$ far away from each other, which makes the average variance of the dimensions greater than the margin $m_d$. In the extreme case, consider each dimension of $\bp_1$ goes to $+\infty$ and each dimension of $\bp_2$ goes to$-\infty$.

The partially collapsed solution is (almost) as bad as the collapsed solution. It assigns the same representation to all the nodes in the same component which makes them inseparable, while the nodes in each component should be separated based on their interactions with the other nodes.

To avoid the partially collapsed solution, we look at the correlation coefficients between the dimensions of the representations. If we have only two unique representations in the user-item representations matrix $\bZ$, then there is a linear relationship between the dimension of the $\bZ$, and by definition the correlation between the dimensions becomes maximum ($-1$ or $+1$). To avoid this, we add an orthogonality term to the objective function to make the representations orthogonal:
%\begin{multline}
%\label{e:obj2}
%E_{ours} = \lambda_1E_{\text{cont}}(\bZ^u,\bZ^i) + \lambda_2E_{d_p}(\bZ) + \lambda_3E_{\text{orth}}(\bZ)  =\\ \lambda_1 \sum_{j,k \in S^+} || \bz^{u}_j - \bz^{i}_k||^2 + \lambda_2 \max{(0,m_p - d_p)}^2 +\\ \lambda_3 \sum_{q=1}^d\sum_{s=q+1}^d \hat{\bZ}_{:,q}^T\hat{\bZ}_{:,s},
%\end{multline}
\begin{equation}
\label{e:obj2}
E_{ours} = \lambda_1E_{\text{cont}}(\bZ^u,\bZ^i) + \lambda_2E_{d_p}(\bZ) + \lambda_3E_{\text{orth}}(\bZ),
\end{equation}
where $E_{\text{cont}}(\bZ^u,\bZ^i)$ and $E_{d_p}(\bZ)$ are defined in Eq.~\eqref{e:obj1} and we defined $E_{\text{orth}}(\bZ)$ as follows:
\begin{equation}
E_{\text{orth}}(\bZ) = \sum_{q=1}^d\sum_{s=q+1}^d \hat{\bZ}_{:,q}^T\hat{\bZ}_{:,s},
\end{equation}
where $\hat{\bZ}$ is achieved by subtracting the mean of each dimension from $\bZ$, and $\hat{\bZ}_{:,q}$ is the $q$th column of $\hat{\bZ}$.
The orthogonality term let the objective avoid the partially collapsed solution by making the off-diagonal values of the covariance matrix small, which consequently makes the correlation of the dimensions of the representations small.

\subsection{Shrinking solution: why we need both terms}
As explained above, if we only optimize the hinge pairwise distance loss $E_{d_p}(\bZ)$ and $E_{\text{cont}}(\bZ^u,\bZ^i)$ together, without using the orthogonality term $E_{\text{orth}}(\bZ)$, it will converge to a partially collapsed solution. On the other hand, optimizing the orthogonality term $E_{\text{orth}}(\bZ)$ and  $E_{\text{cont}}(\bZ^u,\bZ^i)$, without  $E_{d_p}(\bZ)$, is enough to avoid both the collapsed and partially collapsed solutions.

Then, a question arises here: do we need both hinge pairwise distance loss and the orthogonality term in the same objective function? Or is it enough to only enforce representation orthogonality to make the correlation small? In other words, where does the solution converge if we set $\lambda_2=0$ in Eq.~\eqref{e:obj2}?

To answer this question, let us consider a trivial (but not collapsed) solution, where all users' and items' d-dimensional representations are generated from a multivariate uniform distribution. More specifically, each of the $d$ dimensions is drawn from the interval $[-r,r]$, independently from other dimensions. In this scenario, the dimensions are uncorrelated, so the orthogonality term $E_{\text{orth}}(\bZ)$ is almost $0$ in practice. The maximum distance between any two representations is $4dr^2$, which is the upper-bound for $E_{cont}(\bZ^u,\bZ^i)$. So by setting $r$ to an extremely small number (but larger than $0$), we can make $E_{cont}(\bZ^u,\bZ^i)$ extremely small. So this solution, which generates low-quality random representations from a multivariate uniform distribution, can make the objective function of Eq.~\eqref{e:obj2} very small if we don't include the hinge pairwise distance loss. That's because as we make $r$ smaller, the space of the representation shrinks, and the variance of each dimension becomes very small. The hinge pairwise distance loss makes sure that the average variance is larger than a margin and avoids this shrinking solution.

\subsection{Other objective functions in RSs}
\label{s:otherobjs}
In this section, we explore the impact of the proposed terms on the different objective functions. We explain what happens if we change the objective from $E_{\text{cont}}$ to another one, achieved by combining loss functions in Eq.~\eqref{e:rb_loss} and mappings in Eq.~\eqref{e:map}. Is it necessary to use both $E_{d_p}$ and $E_{\text{orth}}$ with the objective functions other than $E_{\text{cont}}$?

Let us explain what happens if we replace $E_{\text{cont}}$ by $E_{\text{MSE}}$, defined in Eq.~\eqref{e:con_reg}. By optimizing $E_{\text{MSE}}$ on the similar pairs, the collapsed solution is achieved by assigning all users and items to any point $\bp \in \bbR^d$ such that $||\bp||_2=1$. Let us assume the user-item graph can be partitioned into $V$ components. The partially collapsed solution is achieved by assigning all representations to any two points $\bp_1 \in \bbR^d$ and $\bp_2 \in \bbR^d$, where $||\bp_1||_2=1$, $||\bp_2||_2=1$, such that all the users and items in the same component get the same representation. 

Unlike the previous case where we used $E_{\text{cont}}$, we do not get the shrinking solution if we draw each dimension of each representation from a multivariate uniform distribution in $[-r,r]$. That's because as we make $r$ smaller, the absolute value of the dimensions of the representations decreases, the dot product of the representations becomes smaller and further away from $1$, and the loss function $E_{\text{MSE}}$ increases. So the loss function $E_{\text{MSE}}$ itself  avoids the shrinking solution.

Since we can prevent the collapsed and partially collapsed solutions using the orthogonality term, without converging to a shrinking solution, we cannot gain much by adding $E_{d_p}$. The results on CiteULike dataset in our experimental results section confirm this.

For other objective functions in the literature, one needs to first verify the existence of the collapsed, partially collapsed, and shrinking solutions and then decide which terms to use to prevent them.

\subsection{Computational complexity and batch-wise training}
Here, we analyze the computational complexity of each term in our objective function in Eq.~\eqref{e:obj2}. The time complexity of $E_{\text{cont}}$ and $E_{\text{orth}}$ are $O(|S^{+}|d)$ and $O((m+n)d)$, respectively. For $E_{d_p}$, if we use Eq.~\eqref{e:d_p}, the complexity is $O((m+n)^2d)$. If we use Eq.~\eqref{e:var} to compute the variance, then the time complexity of computing $E_{d_p}$ will decrease to $O((m+n)d)$, which is linear in the total number of users and items.

In our training, we use neural networks of the previous works to extract user and item representations. Since we use batches to train the model and update the representations, the three terms are computed on a batch and will have a much smaller time complexity, which depends on the number of users and items in the batch.
\section{Experiments}
\label{s:exp}
\subsection{Experimental setup}
\label{ss:setup}
We conduct experiments in both warm-start and item cold-start settings. In the warm-start setting, the interaction feedback of the test users and items is available in training data and no new user or item will be added to the system at test time. In the item cold-start setting, there is no feedback history for the test items and we only have access to their side information. 

In the cold-start setting, we use the following two datasets:
\begin{itemize}
\item \textbf{CiteULike \cite{Wang11}}. In this dataset, each user has a set of saved articles and the goal is to recommend new (cold-start) articles to the users. We use the subset provided by \cite{Volkovs17}, which contains $5\,551$ users, $16\,980$ articles, and $204\, 986$ user-article pairs. The interaction matrix $\bR$ is $99.8$\% sparse and $R_{jk}=1$ means user $j$ saved article $k$. A set of $3\,396$ articles are removed from the training data and used as the test cold-start articles.
\item \textbf{Ichiba10m}. This dataset contains $10$ million purchases from a specific category of the Rakuten Ichiba \footnote{\url{www.rakuten.co.jp}}.
There are around $844\,000$ items, $1.4$M users, and  $2\,000$ cold-start items. The dataset is $99.9991$\% sparse. The model uses the user interaction vector and item side information, which includes item's category, price, and title. The train/test split is done based on the time period. The cold start items are the items that didn't appear during the training period.
\end{itemize}
In Ichiba10m dataset, the goal is to find interested users for a newly released (cold-start) item. In CiteULike dataset, the goal is to assign a set of cold-start articles to each user. In both datasets, we report recall to evaluate the performance of the methods, which shows what portion of the retrieved items/articles is relevant to the users. Since ranking all users is time-consuming in Ichiba, we randomly sample 1 000 unobserved interactions for each item and report recall by returning top 50 users with highest score.

We use the following two datasets in the warm-start setting:
\begin{itemize}
\item \textbf{Lastfm}\footnote{https://www.last.fm/}. The dataset\footnote{http://ocelma.net/MusicRecommendationDataset/lastfm-1K.html} contains $69\,149$ interactions from  $1\,741$ users on $2\,665$ songs. The dataset is $98.5$\% sparse.
\item \textbf{AMusic} \citep{He16a}. The dataset contains $1\,700$ users, $13\,000$ musics, and $46\,000$  interactions. The dataset is $99.8$\% sparse.
\end{itemize}
We use the pre-processed datasets provided by \citet{Dong19}, which is publicly available for non-commercial use\footnote{\url{https://github.com/familyld/DeepCF}}.

We report Hit Ratio (HR) to evaluate the implicit feedback prediction performance. We follow the same protocols as \citep{Dong19, He17} and truncate the rank list at $10$ for both metrics. HR measures whether the actual test item exists in the top-ranked list. 

\paragraph{Baselines.} We compare our method with the state-of-the-art methods in both cold-start and warm-start settings:
\begin{itemize}
\item \textbf{Shared} \citep{Raziperchikolaei21a}. It uses a shared item network to learn both user and item representations. The item network utilizes side information to create item representations. The user representations are generated from the representations of the items purchased by the user.
\item \textbf{Shared attention}  \citep{Raziperchikolaei21a}. Same the the Shared model above, except that it uses attention mechanism to combine the item representations.
\item \textbf{DropoutNet-WMF} and \textbf{DropoutNet-CDL} \citep{Volkovs17}. The key idea is to apply input dropout to the neural network model such that it can handle the missing input data in the cold-start setting.
\item \textbf{ACCM} \citep{Shi18}. It applies the attention mechanism to hybrid RSs to adjust the importance of the input sources.
\item \textbf{BUIR-ID} \citep{Lee21}.  This is the first work to utilize the stop-gradient approach \cite{Chen21} in RSs. Similar to SimPDO, BUIR-ID only uses positive pairs. The difference from SimPDO is that BUIR-ID has two encoders and uses the momentum mechanism to update the parameters of one encoder and to avoid the collapse solution.
 \item \textbf{CFNet} \citep{Dong19}. It uses interaction vector as the input and learns two user and two item representations by two neural network models, which are then combined and mapped to the final interaction by an MLP.
\item \textbf{NeuMF} \citep{He17}. It uses IDs as the input and learns two user and two item representations, which are then combined by an MLP to predict the final interaction.
\item \textbf{DMF} \citep{Xue17}. It learns user and item representations by training a neural networks which uses the explicit feedback as the input and the dot-product as the mapping function.
\end{itemize}

\begin{figure*}
	\begin{center}
			\begin{tabular}{c@{\hspace{1ex}} c@{\hspace{1ex}}c@{}}
			\multicolumn{3}{c}{\dotfill Ichiba10m \dotfill} \\
			\includegraphics[width=0.3\linewidth]{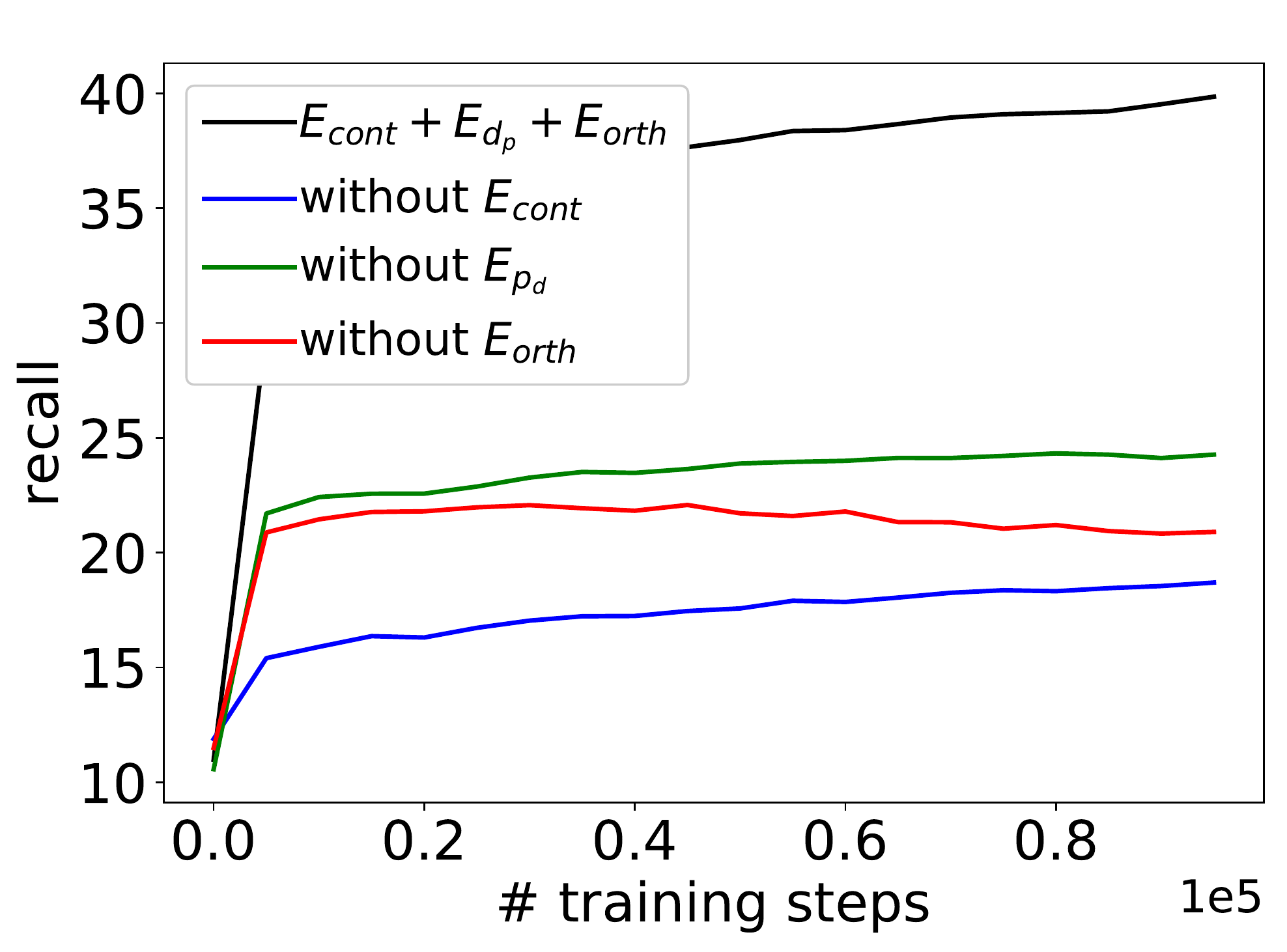}&
			\includegraphics[width=0.3\linewidth]{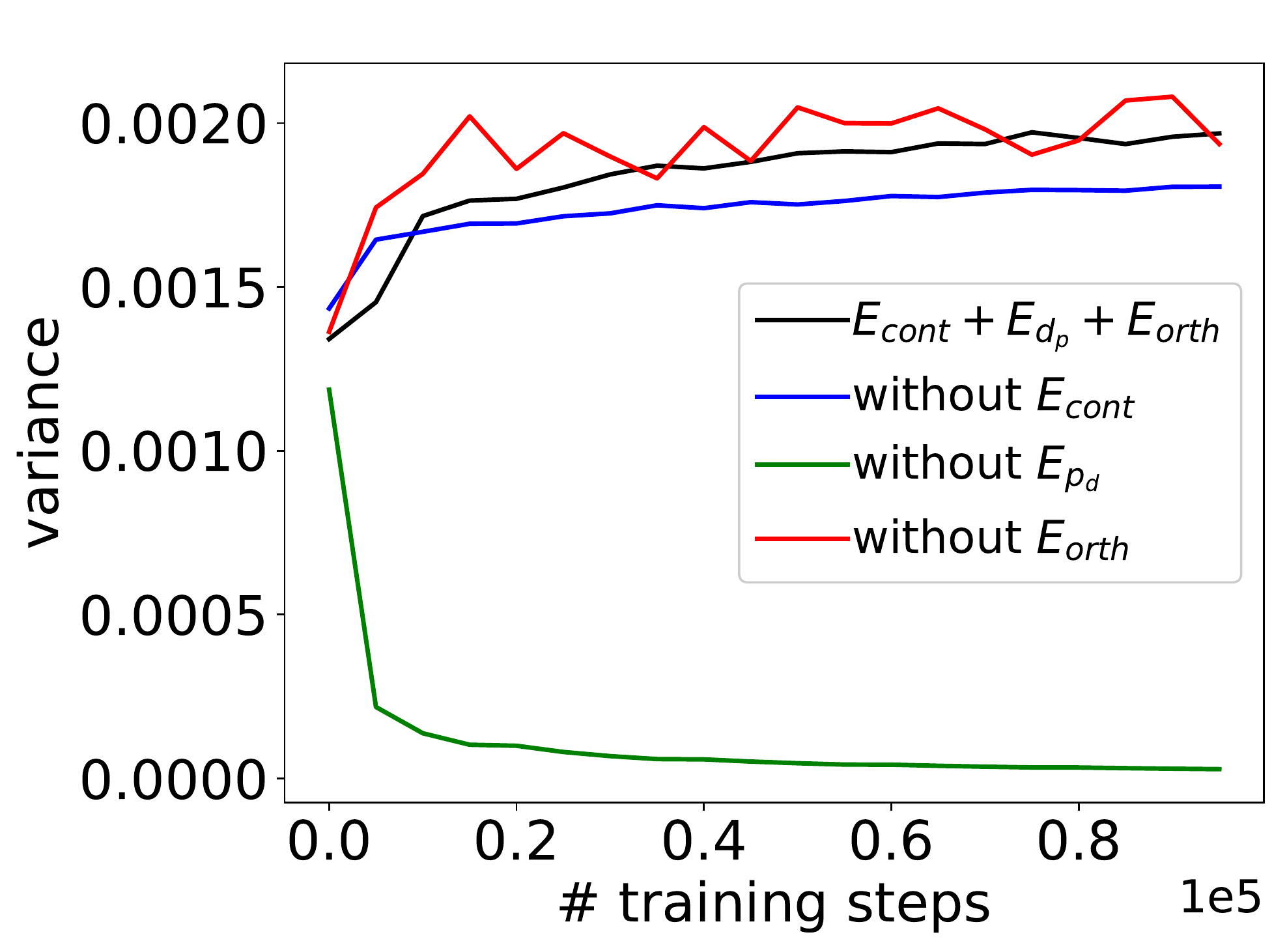} & 				    
			 \includegraphics[width=0.3\linewidth]{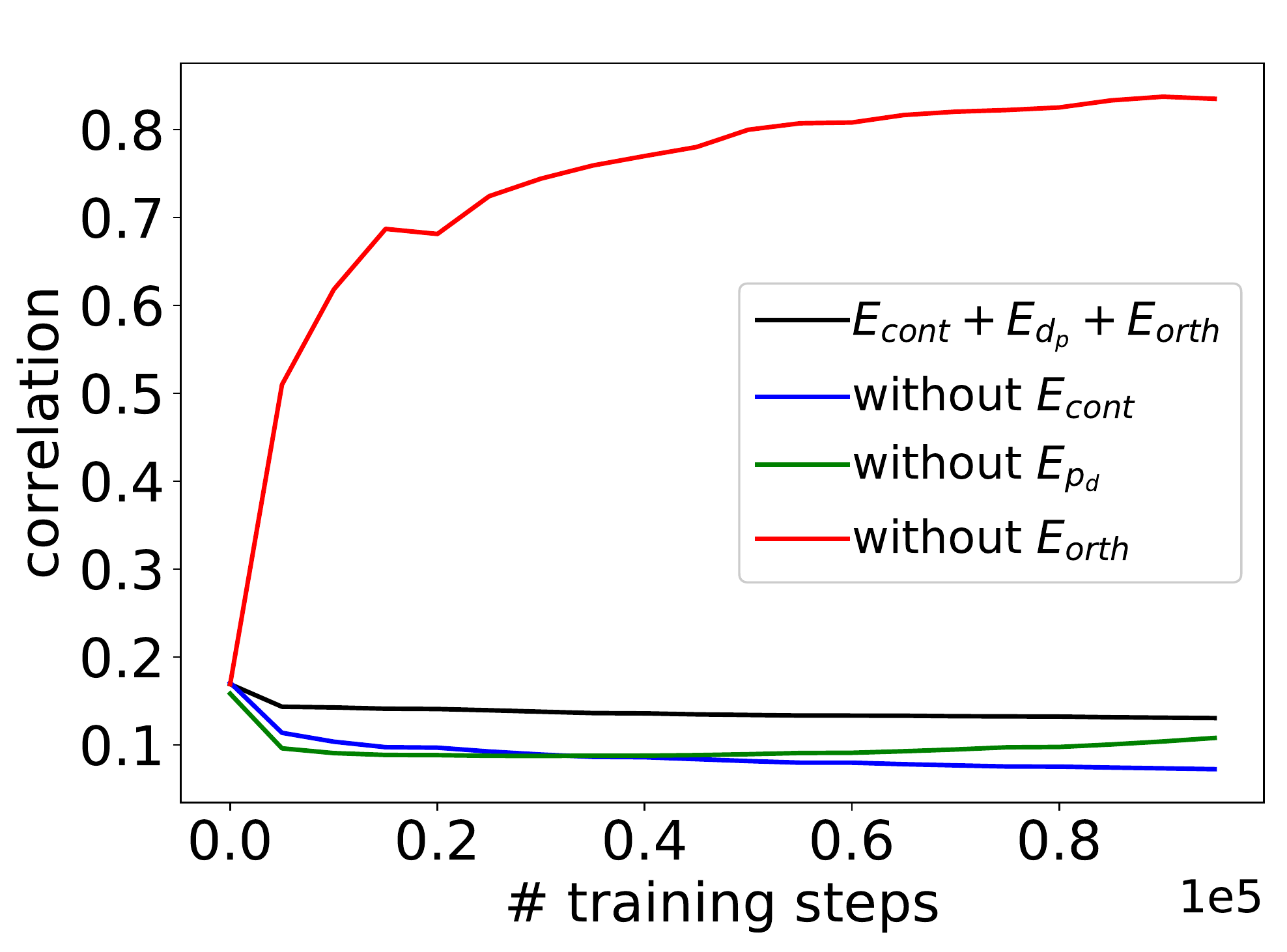}\\
			 \multicolumn{3}{c}{\dotfill CiteULike \dotfill} \\
			 \includegraphics[width=0.3\linewidth]{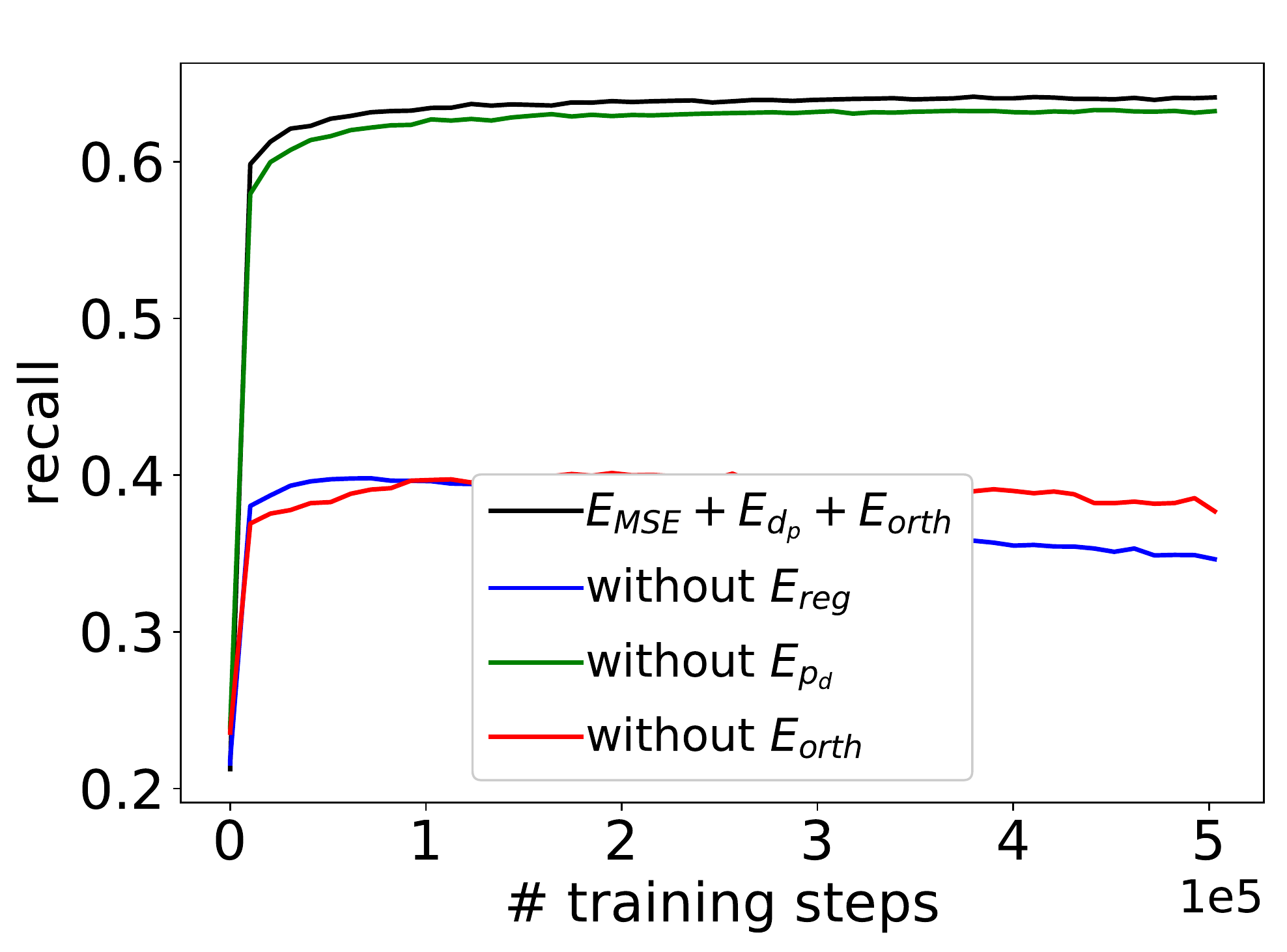} &
			\includegraphics[width=0.3\linewidth]{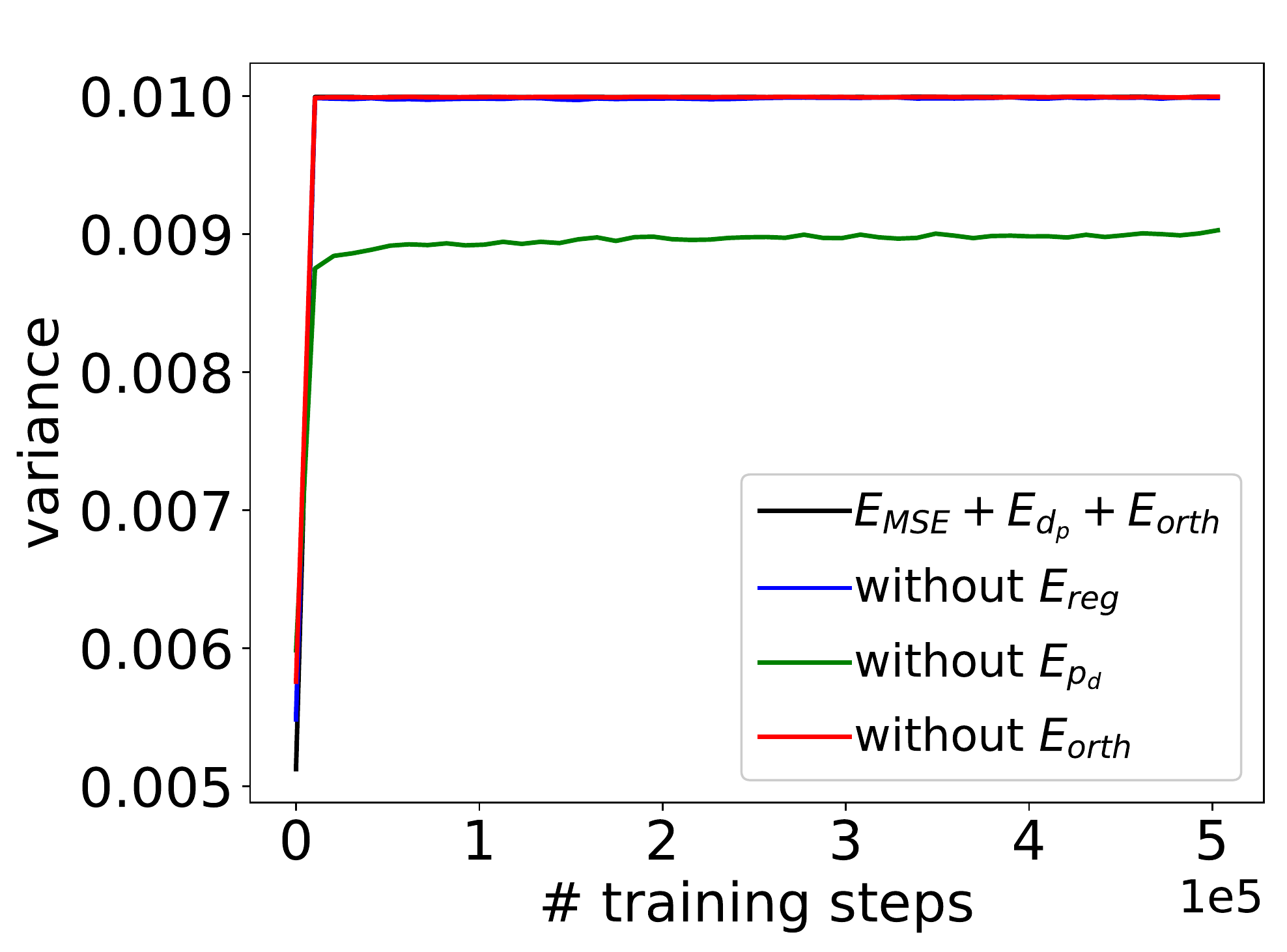} & 			
			 \includegraphics[width=0.3\linewidth]{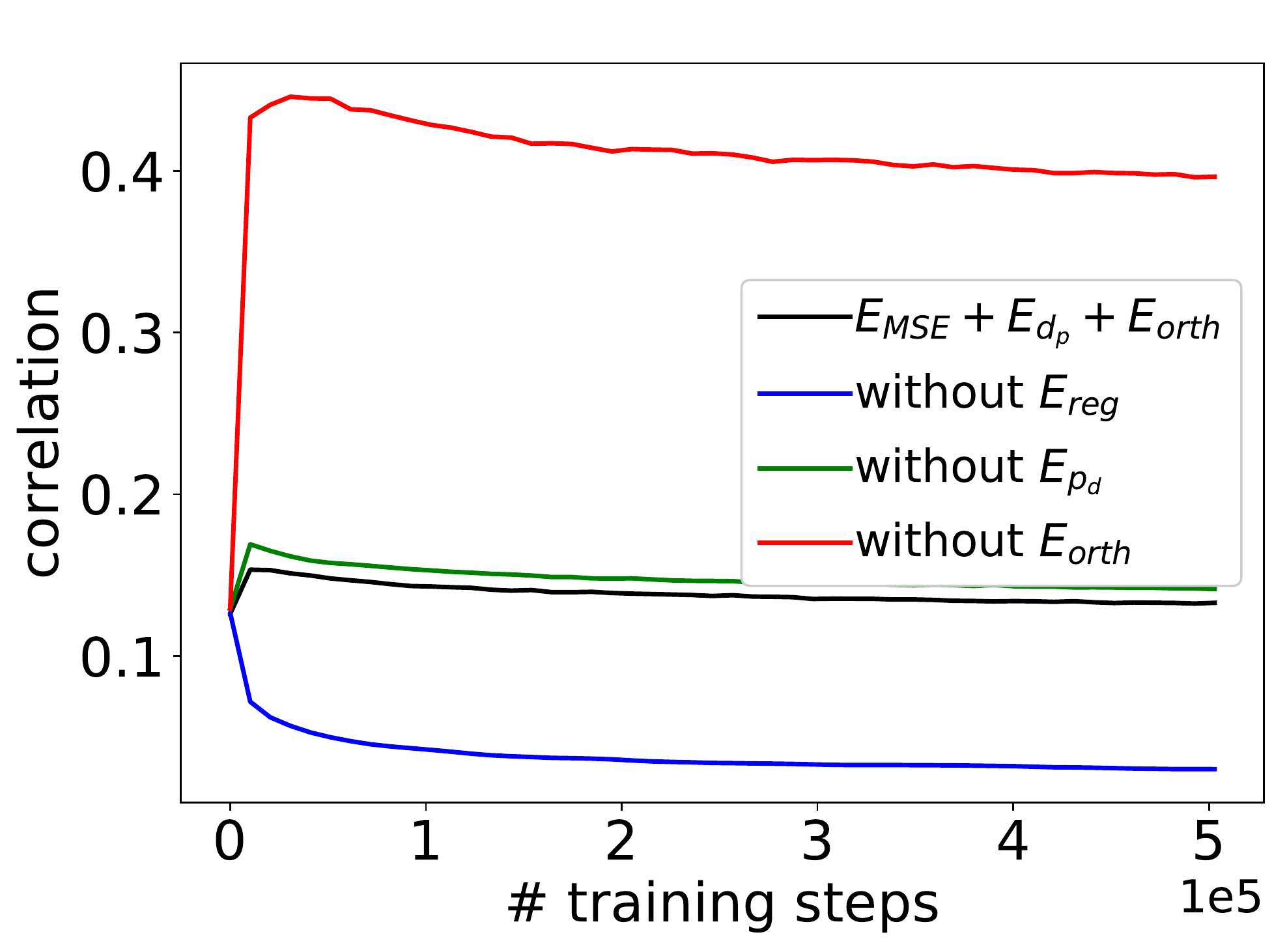} \\
			 \end{tabular}
	\end{center}
	\vspace{-4ex}
\caption{Impact of each term of our objective function on the results in Ichiba10m (top row) and CiteULike (bottom row) datasets. We report the average variance of each dimension, the average correlation between the dimensions, and recall, as we train the models. The model that uses all three terms achieves the best results.}
\label{f:ablation}
\end{figure*}

\paragraph{RSs models to learn the user and item representations.} The specific models to map users and items to the low-dimensional representations are taken from previous works since the novelty of our work is in designing a new objective function, which works no matter what the model is. In the cold-start setting, we use the Shared model proposed by \citet{Raziperchikolaei21a} and in the warm-start setting we use the neural collaborative framework proposed by \citet{Wang19a}.

\paragraph{Implementation details.}
We implemented our method using Keras with TensorFlow 2 backend. We ran all the experiments on one Nvidia Tesla V100-SXM2 32GB GPU in the internal cluster. We use grid search to set the hyper-parameters using a subset of training set and a small validation set. We set the maximum number of epochs to $50$. We use SGD with the learning rate of $0.5$ for our method in all experiments. We set the batch size to $32$ in CiteULike and $128$ in other datasets. We set $\lambda_1=0.01$, $\lambda_1=1$, and $\lambda_3=1$ in all datasets. The margin $m_p$ is $0.1$ in the CiteULike and $0.01$ in all other datasets. The dimension of the user and item embeddings are $100$ in the CiteULike and Ichiba10m dataset and $1\,000$ in AMusic and Lastfm.
\begin{figure}
	\centering
	\includegraphics[width=0.5\linewidth]{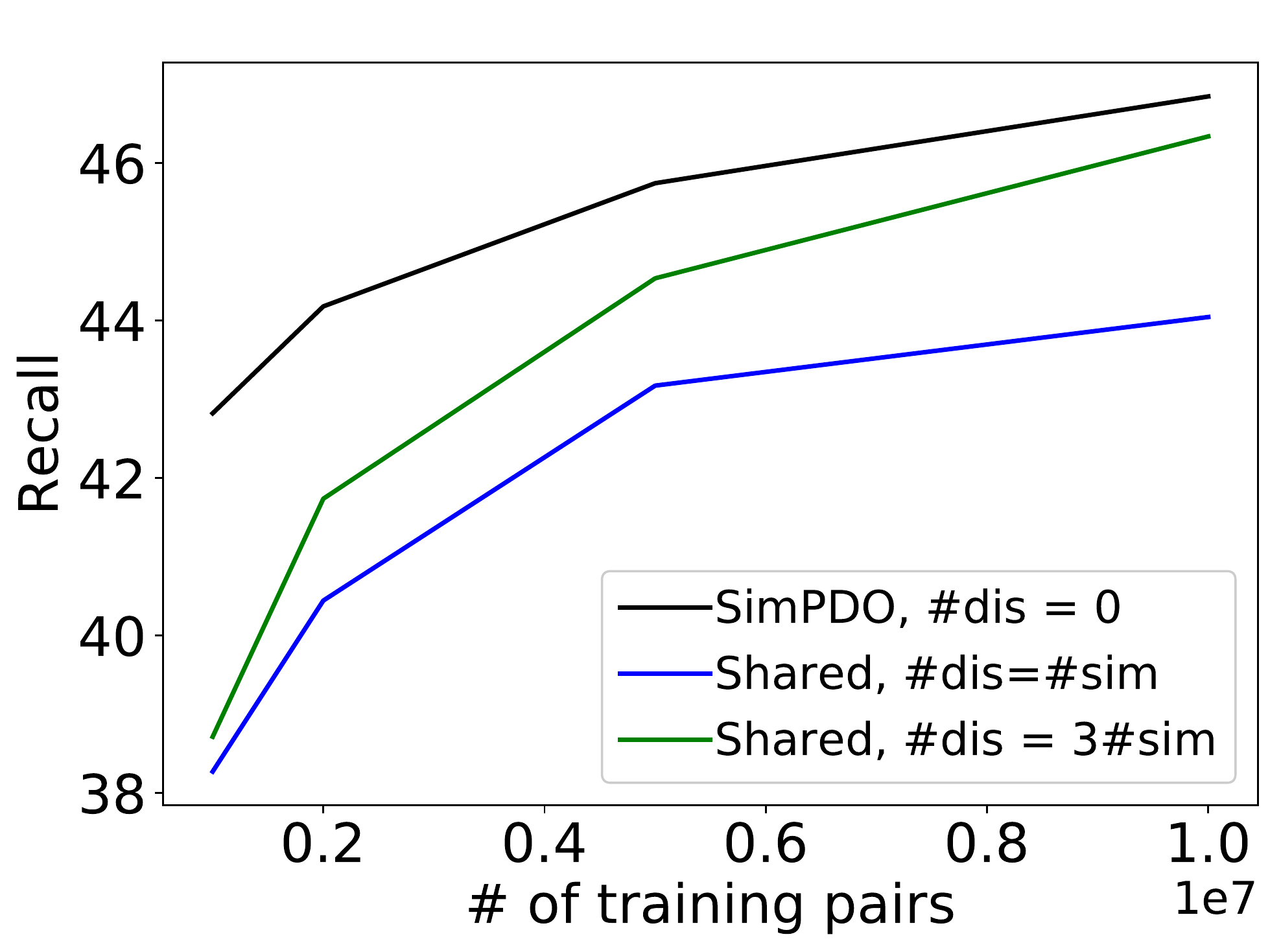}
	\vspace{-2ex}
	\caption{SimPDO vs Shared \cite{Raziperchikolaei21a} as we change the number of training pairs from $1$M to $10$M and report recall@50 in Ichiba10m dataset. SimPDO achieves significantly better results when we use a smaller number of training pairs.}
	\label{f:nump}
\end{figure}

\subsection{Experimental results}
Our proposed objective function uses \textbf{Sim}ilar pairs, \textbf{P}airwise \textbf{D}istance loss, and \textbf{O}rthogonality loss, and is denoted by \textbf{SimPDO}.

\paragraph{Impact of each term of SimPDO.} 
Our proposed objective function has three terms. The first term, $E_{\text{cont}}$ or $E_{\text{MSE}}$, tries to generate a high interaction score for the similar pairs, either by putting their representations close to each other or by maximizing the dot product of the representations. The second term, $E_{d_p}$ defined in Eq.~\eqref{e:d_p}, tries to maximize the pairwise distance of all the pairs (or equivalently maximize the variance) and avoid the collapsed and shrinking solutions. The third term, $E_{\text{orth}}$ in Eq.~\eqref{e:obj2}, avoid the partially collapsed solution by making the representations orthogonal.

In Fig.~\ref{f:ablation}, we investigate the impact of each term by reporting: 1) the average variance of the dimensions of all the representations, 2) the average correlation between the dimensions of the representations, and 3) the recall, as we train the models. 

In the top row, we report the results on Ichiba10m dataset. We can see that the method with all three terms achieves the maximum recall because: 1) without $E_{\text{cont}}$, the similarity between the similar pairs will not be preserved, 2) without $E_{d_p}$, as we can see in the second column, the variance drops significantly towards $0$, which is a sign of the shrinking or collapsed solution, and 3) without $E_{\text{orth}}$, as we can see in the third column, the average correlation increase, which is a sign of the partially collapsed solution.

In the bottom row of the Fig.~\ref{f:ablation}, we show the results on CiteULike, where we use $E_{\text{MSE}}$ instead of the $E_{\text{cont}}$. The results on the CiteULike confirm that the model with all three terms performs best. Note that in CiteULike the curve without $E_{d_p}$ performs almost as well as using all the terms. In addition, note that even without $E_{d_p}$, the variance does not drop significantly and it is a bit smaller compared to the other curves. That's because $E_{\text{orth}}$ avoids the collapsed and partially collapsed solution and the shrinking solution does not happen when we use $E_{\text{MSE}}$.

\paragraph{Training only on similar pairs: better results with fewer training points.}  In Ichiba10m dataset, we fix the test set to $2\,000$ cold-start items, change the number of training pairs from $1$M to $10$M, and report the recall of our method (SimPDO) and Shared model \cite{Raziperchikolaei21a} in Fig.~\ref{f:nump}. SimPDO only uses similar pairs in training and optimizes the objective function of Eq.~\eqref{e:obj2}. Shared model \cite{Raziperchikolaei21a} uses both similar and dissimilar pairs and optimizes the objective function of Eq.~\eqref{e:con}.

\begin{table}
	\caption{Comparison with the state-of-the-art cold-start methods on CiteULike dataset. The * in front of the methods' name means we report the results from the Shared model paper \cite{Raziperchikolaei21a}. Our method outperforms the competitors. We run SimPDO three times and report the mean and standard deviation of the results.}
	\vspace{1ex}
	\label{t:comp_cold}
	\centering
			\begin{tabular}{@{}c@{\hspace{1ex}}c@{}}
	\toprule
	method & recall@100\\
	\midrule
	 \textbf{SimPDO} & $\mathbf{66.8} \pm 0.11$\%\\
	\midrule
	 Shared*  \cite{Raziperchikolaei21a} & $65.7$\%\\
	 \midrule
	 Shared attention* \cite{Raziperchikolaei21a} & $66.4$\%\\
	 \midrule
	 DropoutNet-WMF*  \cite{Volkovs17}& $65.2$\%\\
	 \midrule
	 DropoutNet-CDL* \cite{Volkovs17}& $62.9$\%\\
%	 \midrule
%	 CTR* \cite{Wang11} & $58.9$ \\
%	 \midrule
%	 DeepMusic*  \cite{Oord13} & $60.1$\\
%	 \midrule
%	 CDL* \cite{Wang15b} & $57.3$ \\
	 \midrule
	 ACCM \cite{Shi18} & $63.1$\%\\
	 \bottomrule
	 	\end{tabular}
\end{table}

In SimPDO, all training pairs are similar pairs so we have no dissimilar pairs. In Shared, the training pairs are divided between the similar and dissimilar pairs in different portions. In Fig.~\ref{f:nump}, we considered two cases, where the number of dissimilar pairs is equal to (first case) and three times more than (second case) the number of similar pairs.

There are three remarkable points about the results of Fig.~\ref{f:nump}. First, SimPDO performs better than the Shared model no matter how many training pairs are used, which shows the advantage of training only on similar pairs. Second, the gap between the Shared and SimPDO becomes smaller as we increase the portion of the dissimilar pairs compared to the similar ones, which shows the importance of using a large number of dissimilar pairs. Third, SimPDO is significantly better than the Shared model using a smaller number of training pairs. This is a big advantage of SimPDO when the datasets have billions of pairs and training on all of them is time-consuming: SimPDO can be trained on a smaller training set and still achieve reasonable results.

\begin{table}
\caption{Comparison with the state-of-the-art warm-start methods on AMusic and Lastfm datasets. The * in front of the methods' name means we report the results from the DeepCF paper \cite{Dong19}. Our method achieves competitive results using smaller number of training pairs. We put $-$ for HR@5 of some of the methods since only HR@10 is reported in \cite{Dong19}. We run SimPDO, BUIR-ID, and CFNet three times and report the mean and standard deviation of the results.}
\vspace{1ex}
\label{t:comp_warm}
		\centering
		\begin{tabular}{@{}c@{\hspace{.3ex}}c@{\hspace{1ex}}c@{\hspace{1.5ex}}c@{\hspace{1ex}}c@{}}
	\toprule
	& \multicolumn{2}{c}{AMusic} & \multicolumn{2}{c}{Lastfm}\\
	method & HR@5  & HR@10  & HR@5 & HR@10\\
	\midrule
	 \textbf{SimPDO}  & $\mathbf{32.66 \pm 0.4}$ &  $\mathbf{43.47 \pm 0.09}$  & $75.4 \pm 0.1$  & $\mathbf{89.6 \pm 0.18}$  \\
	 \midrule
	 BUIR-ID \citep{Lee21} & $25.76 \pm 0.2$ & $36.5 \pm 0.7$ & $70.16 \pm 0.6$ & $84.15 \pm 0.9$     \\
	 \midrule
	 CFNet \citep{Dong19} & $29.7 \pm 0.08$ & $38.4 \pm 0.003$& $\mathbf{77.1 \pm 0.17}$ & $88.6 \pm 0.02$    \\
	\midrule
	 CFNet* \citep{Dong19} & - & $41.2$ & - &$89.9$    \\
	 \midrule
	 CFNet-ml* \citep{Dong19}  & - & $40.7$  &- &$88.3$  \\
	 \midrule
	 CFNet-rl* \citep{Dong19} & - & $39.4$   &- &$88.4$  \\
	 \midrule
	 NeuMF* \citep{He17} & - & $38.9$   &- &$88.6$ \\
	 \midrule
	 DMF* \citep{Xue17} & - & $37.4$  &- &$88.4$ \\
	 %\midrule
	 %eALS* \shortcite{He16b}  & - & $37.1$  &- &$82.6$ \\
	 \bottomrule
	 	\end{tabular}
\end{table}

\begin{table}
\caption{We report number of similar and dissimilar pairs to train different methods and training time per epoch of different methods. SimPDO and BUIR-ID use a smaller number of pairs and can be trained much faster than the other methods.}
\vspace{1ex}
	\label{t:tt}
		\centering
		\begin{tabular}{@{}c@{\hspace{.3ex}}c@{\hspace{1ex}}c@{\hspace{1ex}}c@{\hspace{1.5ex}}c@{\hspace{1ex}}c@{\hspace{1ex}}c@{\hspace{1.5ex}}c@{\hspace{1ex}}c@{\hspace{1ex}}c@{}}
	\toprule
	& \multicolumn{3}{c}{\dotfill AMusic \dotfill } & \multicolumn{3}{c}{\dotfill  Lastfm \dotfill } & \multicolumn{3}{c}{\dotfill CiteULike \dotfill }\\
	method & numS  & numD & training time & numS & numD&  training time& numS & numD&  training time\\
	\midrule
	 \textbf{SimPDO}  & $46$K &$0$ & $13$s & $69$K & $0$ & $15$s& $200$K & $0$M & $123$s  \\
	 \midrule
	 \textbf{BUIR-ID}  & $46$K &$0$ & $8$s & $69$K & $0$ & $11$s & $200$K & $0$M & $123$s  \\
	 \midrule
	 CFNet & $46$K& $184$K & $36$s & $69$K & $276$K & $36$s & - & -& -  \\
	 \midrule
	 Shared & -& -  & - & -& -&-& $200$K & $2$M & $ 1871$s\\
	 \midrule
	 Shared\_attention  & - & - & - &-&-&-&$200$K & $2$M & $ 1963$s\\
	 \bottomrule
	 	\end{tabular}
\end{table}

\paragraph{Comparison with the state-of-the-art methods.}
In Table~\ref{t:comp_cold}, we compare the methods on CiteULike dataset in the cold-start setting.  The dataset and the Shared code are available online\footnote{\url{https://github.com/rakutentech/shared_rep}}. We use the code and processed data provided in this repository to train our model. Note that our method uses the same user and item models as the shared model, but uses the new objective function. This makes it clear that the better performance of the SimPDO comes from the proposed objective function that is only using similar pairs.

In Table~\ref{t:comp_warm}, we compare the methods on AMusic and Lastfm datasets in the warm-start setting.  The dataset and the DeepCF code are available online\footnote{\url{https://github.com/familyld/DeepCF}}. We use the code and processed data provided in this repository to train our model. We also trained CFNet and BUIR-ID from the publicly available code and put the results in the second and third row of Table~\ref{t:comp_warm}.

Results of Table~\ref{t:comp_cold} and Table~\ref{t:comp_warm} show that SimPDO outperforms most state-of-the-art methods and achieves very competitive results. Note that SimPDO and BIUR use a significantly smaller number of training pairs compared to the other methods, such as Shared and CFNet. We have put the number of similar and dissimilar pairs used in training the methods with the best results in Table~\ref{t:tt}. SimPDO and BUIR-ID use a smaller number of pairs and can be trained much faster than the other works.

\section{Conclusion}
In this paper, we proposed SimPDO, a new objective function that enables training of one-class recommendation system models without dissimilar pairs. We showed that by only using similar pairs, the optimal solution of existing objective functions becomes a collapsed solution, where every representation is mapped to the same point in the latent space. We avoided the collapsed solution by providing a hinge loss for pairwise distance which is equivalent to the average variance of each dimension of representations. We also showed that we need an orthogonality term to avoid collapsed and partially collapsed solutions, where the optimal solution under the pairwise loss returns only two representations. The orthogonality term minimizes the correlation between each dimension of representations and forces the objective function to return various representations. Finally, we showed that both terms are necessary to learn meaningful representations.
The results demonstrated that SimPDO outperformed the existing RS objective functions without using dissimilar pairs. We also showed that SimPDO can be trained more efficiently with a smaller number of training pairs.

{\small
\bibliographystyle{plainnat}
\bibliography{lib_rakut}
}

\end{document}